\documentclass[twocolumn]{aastex63}

\accepted{to ApJ 31 October 2022}

\newcommand{\cha}{\textit{Chandra }}
\def\xmm{{XMM-{\it Newton\/}}}
\def\fermi{{{\it Fermi}-LAT }}

\def\aap{A\&A}

\def\apjl{ApJL}
\def\apjs{ApJS}

\shorttitle{MeV--GeV $\gamma$-ray Emission from SNR~G327.1--1.1 Discovered by the \fermi}

\shortauthors{Eagle et al.}

\PassOptionsToPackage{colorlinks,linkcolor={blue},citecolor={blue},urlcolor={red}}{hyperref}
\usepackage{hyperref}
\usepackage{natbib}
\usepackage{float}
\usepackage{color}
\usepackage{graphicx}
\usepackage{gensymb}
\usepackage{amsmath}
\usepackage{enumitem}
\usepackage{multirow,helvet}
\usepackage{todonotes}
\usepackage{lineno}
\usepackage{lipsum}

\begin{document}



\title{MeV--GeV Gamma-ray Emission from SNR~G327.1--1.1 Discovered by the \fermi}

\correspondingauthor{Jordan Eagle}
\email{jeagle@clemson.edu}

\author[0000-0001-9633-3165]{Jordan Eagle}
\affiliation{Harvard-Smithsonian Center for Astrophysics\\
Cambridge, MA 02138, USA}
\affiliation{Department of Physics \& Astronomy\\ 
Clemson University\\ 
Clemson, SC 29634, USA}

\author[0000-0002-0394-3173]{Daniel Castro}
\affiliation{Harvard-Smithsonian Center for Astrophysics\\
Cambridge, MA 02138, USA}

\author[0000-0001-7380-3144]{Tea Temim}
\affiliation{Department of Astrophysical Sciences\\
Princeton University\\ 
Princeton, NJ, 08544, USA\\}

\author[0000-0002-8784-2977]{Jean Ballet}
\affiliation{Universit\'{e} Paris-Saclay, Universit\'{e} Paris Cit\'{e}, CEA, CNRS, AIM, 91191 Gif-sur-Yvette, France \\}

\author[0000-0002-6986-6756]{Patrick Slane}
\affiliation{Harvard-Smithsonian Center for Astrophysics\\
Cambridge, MA 02138, USA}

\author[0000-0003-4679-1058]{Joseph Gelfand}
\affiliation{New York University Abu Dhabi, P.O. Box 129188, Abu Dhabi, United Arab Emirates\\}

\author[0000-0002-0893-4073]{Matthew Kerr}
\affiliation{Space Science Division, Naval Research Laboratory, Washington, DC 20375-5352, USA\\ }

\author[0000-0002-6584-1703]{Marco Ajello}
\affiliation{Department of Physics \& Astronomy\\ 
Clemson University\\ 
Clemson, SC 29634, USA}

\begin{abstract}
We report the discovery of MeV--GeV $\gamma$-ray emission by the \fermi positionally coincident with the TeV pulsar wind nebula (PWN) HESS~J1554--550 within the host supernova remnant (SNR) G327.1--1.1. The $\gamma$-ray emission is point-like and faint but significant ($> 4 \sigma$) in the 300\,MeV--2\,TeV energy range. 
We report here the \fermi analysis of the observed $\gamma$-ray emission followed by a detailed multiwavelength investigation to understand the nature of the emission. The central pulsar powering the PWN within G327.1--1.1 has not been detected in any waveband; however, it is likely embedded within the X-ray nebula, which is displaced from the center of the radio nebula. The $\gamma$-ray emission is faint and therefore a pulsation search to determine if the pulsar may be contributing is not feasible.  Prior detailed multiwavelength reports revealed an SNR system that is old, $\tau \sim 18,000$\,yrs, where the interaction of the reverse shock with the PWN is underway or has recently occurred. We find that the $\gamma$-ray emission agrees remarkably well with a detailed broadband model constructed in a prior report based on independent hydrodynamical and semi-analytic simulations of an evolved PWN. We further investigate the physical implications of the model for the PWN evolutionary stage incorporating the new \fermi data and attempt to model the distinct particle components based on a spatial separation analysis of the displaced PWN counterparts.
\end{abstract}

\section{Introduction}

Pulsar wind nebulae (PWNe) are descendants of core collapse supernovae (CC SNe), and are powered by an energetic, rapidly rotating neutron star. As the neutron star loses spin, rotational energy is transferred into an ultra-relativistic particle wind, made up of mostly electrons and positrons. A shockwave, called the \textit{termination shock}, is generated within the nebula 
where the ram pressure of the pulsar wind is balanced by the pressure of the highly relativistic plasma 
\citep{pwn2006,slane2017}. The termination shock is the location where particles injected by the central pulsar are accelerated as they enter the non-thermal pool of relativistic particles in the nebula. Synchrotron emission from the relativistic electrons is observed from most PWNe, from radio wavelengths to hard X-rays, while the same relativistic electrons are thought to scatter off of local photon fields, resulting in Inverse Compton (IC) emission at $\gamma$-ray energies. 
Most PWNe have been discovered at radio or X-ray bands with an increasing number of discoveries in TeV. Indeed, the majority of the TeV Galactic source population is found to originate from PWNe as observed by Imaging Air Cherenkov Telescopes \citep[IACTs, e.g.][]{acero2013}. 

The PWN broadband spectrum depends both on the particle spectrum that was initially injected at the termination shock by the pulsar and how it was altered throughout the evolution of the PWN inside its SNR \citep{reynolds_1984, gelfand2009}. The evolution of a PWN is connected to the evolution of the central pulsar, host supernova remnant (SNR), and the structure of the surrounding interstellar medium \citep[ISM,][]{pwn2006}. Eventually, the relativistic particle population will be injected into the ISM to contribute to the Galactic cosmic ray (CR) electron--positron population. Current predictions from semi-analytical models all indicate this occurs in the late-phase of PWN evolution \citep{blondin2001, gelfand2009, temim2015}. Several detailed evolutionary models \citep[e.g.,][]{blondin2001,swaluw2004, gelfand2009} predict the spatial emission properties and dynamical and radiative evolution for PWNe inside a non-radiative SNR using magnetohydrodynamical (MHD) and semi-analytic modeling methods. There has been significant improvement in model performance when considering the pressure produced by both the particles and the magnetic field of the PWN, which influences the compression evolutionary phase that is initiated by the return of the reverse shock \citep[e.g.,][]{bandiera2020}.  Additionally, the PWN and SNR shock structures can be accurately described when accounting for the dependence between the properties of the host SNR and of the PWN  \citep[e.g.,][]{bandiera2021}. Other recent works have incorporated multi-dimensional, time-dependent leptonic transport codes that simulate PWN evolution from birth to present day \citep{van2020}.
At later evolutionary stages as is observed with G327.1--1.1, the interaction between the PWN and the reverse shock is a critical stage that initiates rapid synchrotron losses through the compression of the PWN. The relativistic particle population is significantly disturbed at this stage.  Constraining the physical properties of the particles is difficult, requiring a combination of simulation tools and observable properties over the entire electromagnetic spectrum.  

In this paper we report the detection of faint $\gamma$-ray emission coincident with the middle-aged supernova remnant SNR~G327.1--1.1. In Section \ref{sec:select} we describe the SNR~G327.1--1.1 system. A multiwavelength analysis is described in Section~\ref{sec:mw}, including re-analyzing archival \cha observations to spatially separate the X-ray nebula components and their X-ray spectra in Section \ref{sec:xray}. We present the \fermi data analysis in Section \ref{sec:fermethod}. Broadband modeling incorporating the spatially separated multiwavelength data and the resulting best-fit spectral energy distribution (SED) model is described in Section \ref{sec:model}. We discuss implications of observations and modeling in Section \ref{sec:discuss} and present our final conclusions in Section \ref{sec:conclude}.

\section{SNR~G327.1--1.1}\label{sec:select}
\begin{figure}
\centering
\includegraphics[width=1.0\linewidth]{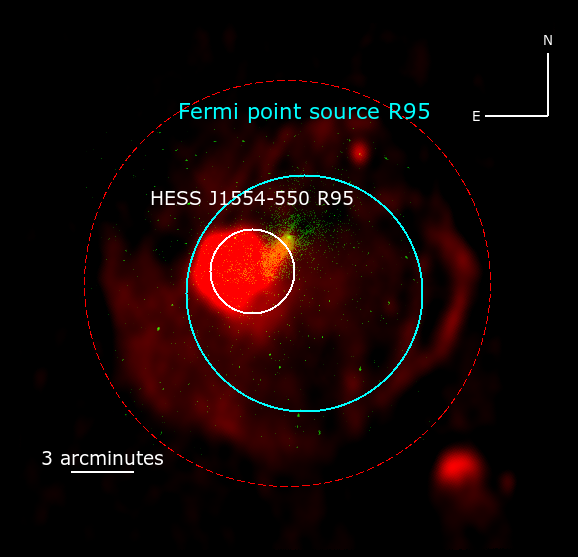}
\caption{SNR~G327.1--1.1 as observed by the Molonglo Observatory Synthesis Telescope (MOST) at 843\,MHz (red), \cha X-rays between 0.5--7\,keV (green), and the H.E.S.S. TeV and \fermi best-fit positions indicated by their 95\% uncertainty radii, $R_{95} = 0.035 \degree$ and $R_{95} = 0.079 \degree$ for the TeV and MeV--GeV positions, respectively. The faint radio SNR shell is outlined as the red-dashed circle.}\label{fig:rgb}
\end{figure}

SNR~G327.1--1.1 is a relatively old remnant with an estimated age $\tau \sim 18,000$\,yrs based on the Sedov--Taylor solution for an evolved SNR and a distance of $9$\,kpc \citep{sun1999,temim2009}. The SNR is visible as a faint radio shell at 843\,MHz accompanied by a bright, asymmetric PWN, see Fig. \ref{fig:rgb}. A similar structure is observed in the X-rays with thermal emission ($kT \sim 0.3$\,keV) present from the SNR shell and strong nonthermal emission (photon index $\Gamma_X \sim 2.1$) from the PWN \citep{sun1999,bepposax2003, temim2009, temim2015}. Radio and X-ray observations suggest the reverse shock, generated from the pressurized ISM gas swept up from the SNR forward shock, has recently interacted with the PWN asymmetrically, compressing and displacing the PWN to the East. As displayed in Figure \ref{fig:rgb}, the radio nebula is displaced from the X-ray emission while the X-ray emission has a cometary morphology, suggesting the pulsar is the point source embedded within the X-ray emission peak. \citet{temim2009} searched for X-ray pulsations within the \cha data, but the PWN outshines the pulsar, accounting for 94\% of the \cha flux for the compact region of the X-ray nebula. The pulsar is likely moving at a high velocity ($\sim 400$\,km s$^{-1}$) to the North based on its location with respect to the geometric center of the SNR. The pulsar's velocity in combination with an asymmetric reverse shock interaction could explain the Southeastern displacement of the PWN \citep{temim2009, temim2015}. This scenario is consistent with the morphology, since the X-ray cometary structure is pointed Northwest and elongates to the Southeast where it meets the radio relic PWN which is displaced to the East. 

It is possible that the formation of the PWN is from the pulsar's high velocity, creating a bow shock with a comet shaped X-ray nebula. However, the presence of the X-ray prongs ahead of the pulsar, in addition to evidence for a torus-like structure embedded inside the cometary nebula, suggests the reverse shock is responsible for the observed morphology. 

\citet{acero2011} reported the first detection of point-like TeV $\gamma$-ray emission in the direction of SNR~G327.1--1.1 observed by the High Energy Stereoscopic System (H.E.S.S.). The TeV source, HESS~J1554--550, is uniquely positioned inside the radio SNR shell and overlaps with the radio and X-ray PWN locations, with an uncertainty radius $R < 0.035$\,$\degree$ (see Figure \ref{fig:rgb}), confirming the high-energy nature of the PWN. MeV--GeV $\gamma$-ray emission from the PWN was recently detected by the \fermi \citep{xiang2021}.

\section{Multiwavelength Information}\label{sec:mw}
\subsection{Radio}
The SNR~G327.1--1.1 is well-known for its composite radio structure which was first revealed in the MOST radio survey \citep{whiteoak_green1996}. The SNR exhibits a faint non-thermal radio shell from the SNR and a bright core of emission with a bulge extending out of the core in the northern direction sometimes referred to as the ``finger." The bulge of radio emission also outlines the X-ray cometary structure and presumed pulsar, supporting the idea that the SNR reverse shock (RS) has crushed the PWN and the pulsar has begun exiting the nebula. 

More recently, \citet{ma2016} reported Australia Telescope Compact Array (ATCA) radio observations at 6 and 3 \,cm that resolves the complex structure for the PWN into two distinct components: a radio ``body'' which is the relic PWN and the radio ``head'' which is the radio ``finger'' structure that overlaps with the compact X-ray nebula \citep[see Fig.~2 and Table~2 of][]{ma2016}. The radio ``body'' or relic component dominates the overall radio spectrum, which is consistent with the picture that the radio relic contains the oldest particles, whereas the radio ``head'' represents a younger nebula with fainter radio emission, but brighter X-ray emission. The total radio spectrum is typical for other observed PWNe with a spectral index $\alpha = -0.3 \pm 0.1$. Additionally, the PWN is found to be highly linearly polarized with polarization fractions between 15--20\% at 6 and 3\,cm. The implied magnetic field direction is aligned along the ``finger'', with at least a partially azimuthal magnetic field structure in the relic nebula.  


\subsection{X-ray}\label{sec:xray}
SNR~G327.1--1.1 has been studied in detail in the X-ray band by the {\it Einstein} Observatory, ROSAT, ASCA, BeppoSAX, \xmm, and \cha \citep[see e.g.,][]{lamb1981, seward1996, sun1999, bepposax2003, temim2009, temim2015}. In Figures~\ref{fig:rgb} and \ref{fig:xray_regions}, the PWN/SNR in X-rays as seen by \cha is displayed between energies 0.5--7\,keV. The X-ray morphology consists of the diffuse SNR thermal emission which overlaps with the fainter SNR shape in radio. A slender nonthermal component in the X-ray, which overlaps well with the radio ``finger,'' corresponds to the PWN and the presumed pulsar, which is most likely the point source embedded within the bright X-ray nebula. Two prong-like structures extend from the Northwest of the X-ray peak. It has been suggested that these structures result from an interaction between the PWN and re-heated SN ejecta, though no thermal emission has been detected \citep{temim2009,temim2015}. While the origin of the X-ray prongs remain unclear, it could alternatively pinpoint to enhancements in the magnetic field \citep{temim2015}. In the next section, we extract and measure the X-ray spectra corresponding to the radio ``body'' or relic and radio ``head'' or finger components.

\subsubsection{\cha Data Analysis}
We re-analyze deep archival X-ray observations of the SNR~G327.1-1.1 performed with the Advanced CCD Imaging Spectrometer, ACIS-I, on board the \cha X-ray Observatory for a total exposure time of 350 ks.  The standard data reduction and cleaning were performed using the Chandra Interactive Analysis of Observations \citep[CIAO v.4.12,][]{ciao2006} software package, resulting in a total clean exposure time of 337.5 ks.  An identical data reduction procedure has already been performed and reported \citep[see][for details]{temim2015}. We re-analyze the X-ray data incorporating two different source extraction regions, following the source regions as those used to measure the distinct radio ``finger'' and ``body'' PWN components from \citet{ma2016}. Figure~\ref{fig:xray_regions} shows the regions used for the spectral extraction. A spectrum for each component is extracted using the \texttt{specextract} tool in CIAO and modeled using SHERPA within CIAO \citep{sherpa2001}. We measure two X-ray source spectra corresponding to the spatially separated radio components and their radio spectra. 

\subsubsection{Chandra Data Analysis Results}
\begin{figure}
\hspace{-0.95cm}
\includegraphics[width=1.2\linewidth]{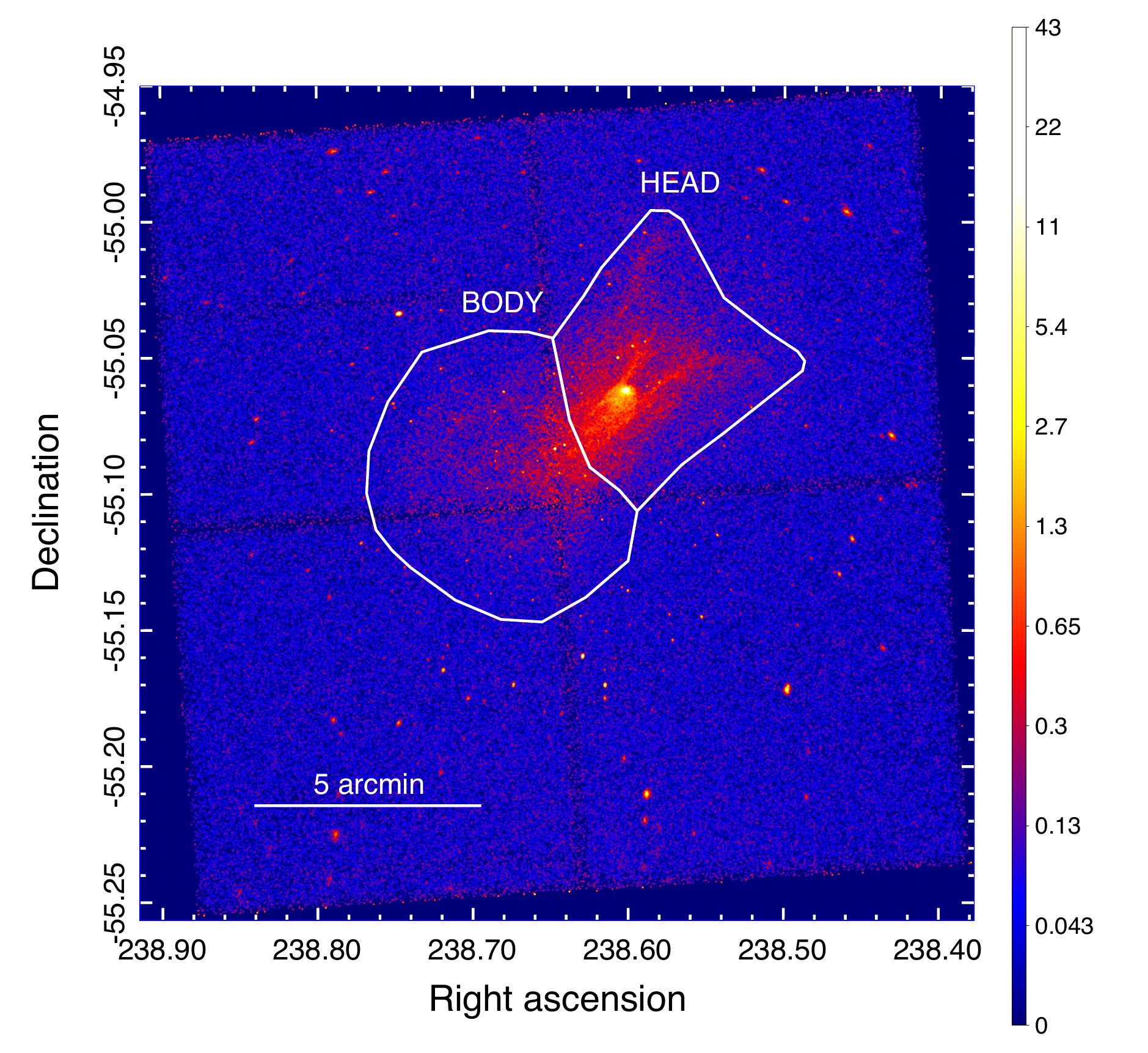}
\caption{The unsmoothed Chandra ACIS-I image of the PWN region (2.6$^\prime$ across) with the ``body'' and ``head'' regions used for the spectral extraction, which are chosen to be the same regions of the radio relic (``body'') and radio finger (``head'') spectral extraction regions used in \citet[][see their Fig.~2, bottom right panel]{ma2016}.}\label{fig:xray_regions}
\vspace{0.05cm}
\end{figure}

\begin{figure*}
\begin{minipage}{0.5\textwidth}
\centering
\includegraphics[width=1.\linewidth]{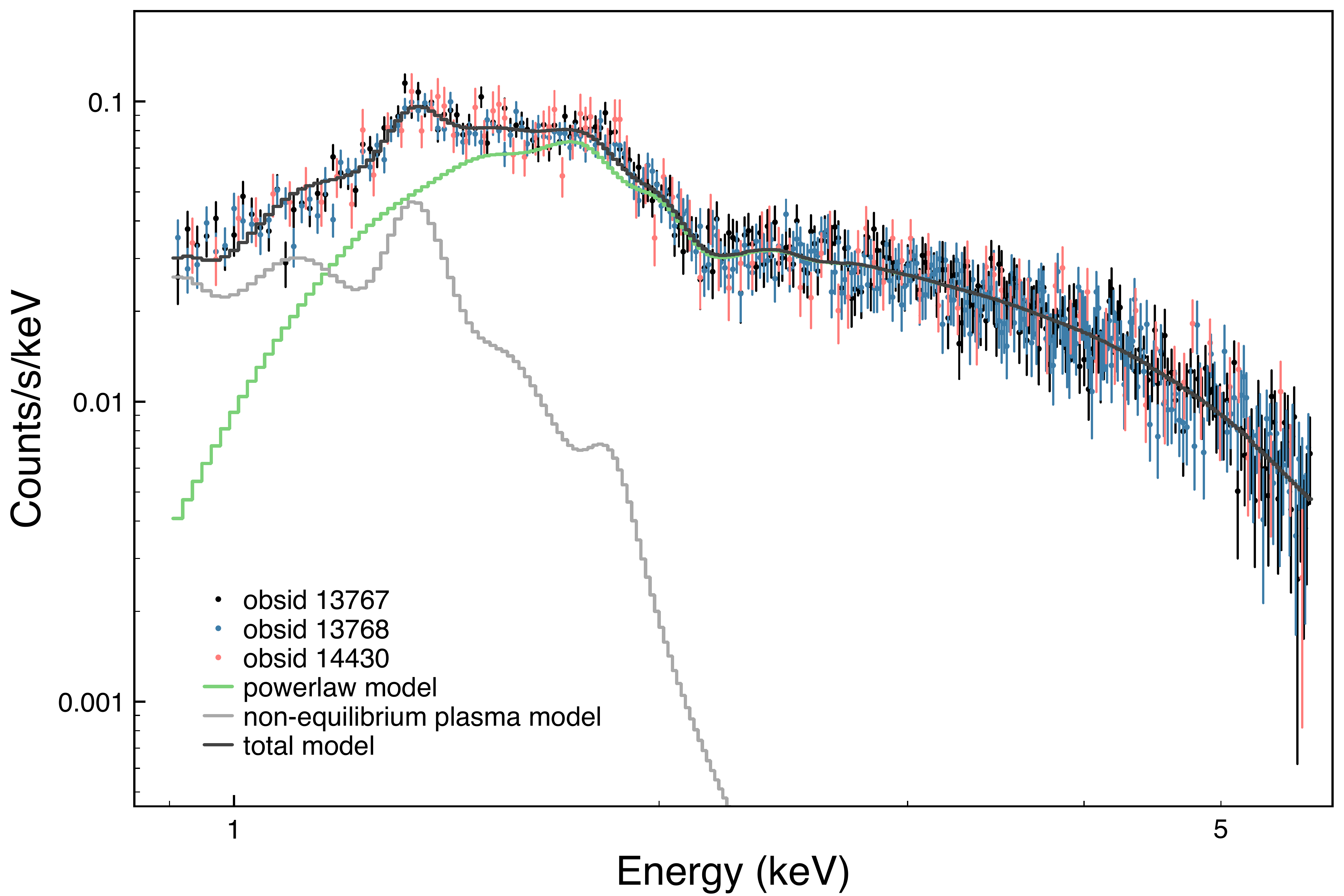}
\end{minipage}
\begin{minipage}{0.5\textwidth}
\centering
\includegraphics[width=1.\linewidth]{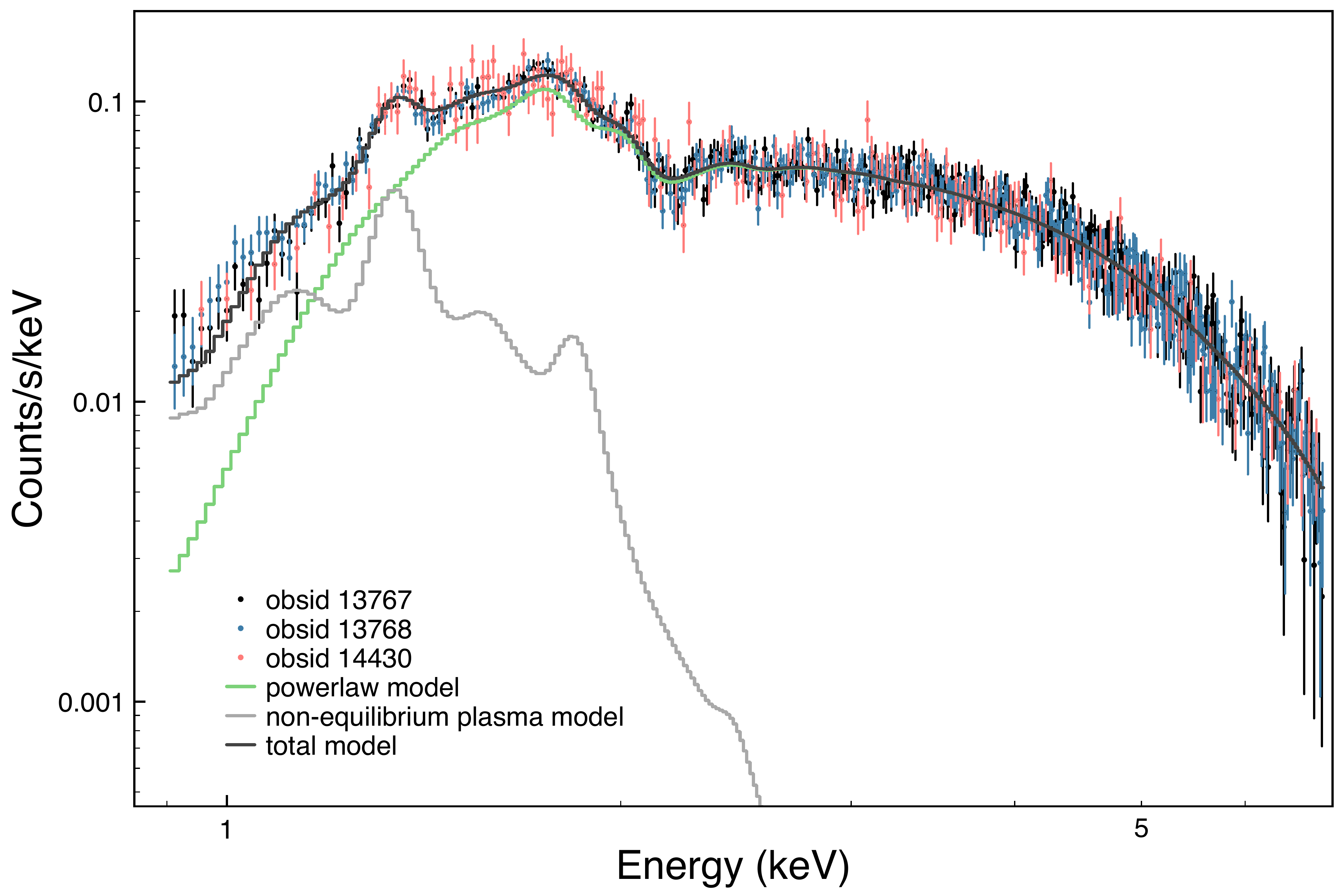}
\end{minipage}
\caption{ {\it Left:} The data and best-fit model for the ``body'' PWN component illustrated in Figure~\ref{fig:xray_regions}. {\it Right:}  The data and best-fit model for the ``head'' PWN component illustrated in Figure~\ref{fig:xray_regions}. In both panels, the nonthermal spectrum originating from the PWN is plotted in green and the thermal spectrum originating from the SNR is plotted in gray. The total X-ray spectral model is plotted in black. }\label{fig:xray_spectra}
\vspace{0.05cm}
\end{figure*}

In order to study the properties of the two distinct structures observed for the PWN, we fitted spectra from two different source regions, shown in Figure~\ref{fig:xray_regions}. Since the SNR shell covers most of the field of view, we used the ACIS blank-sky background files adopted to our observation\footnote{\url{http://cxc.harvard.edu/ciao/threads/acisbackground/}} to extract background spectra from each extraction region. We used the high-energy data from 10 to 12\,keV to compare the particle background in each ACIS-I chip in our observations to the particle background in the blank-sky data. This procedure is identical to the one performed and outlined in \citet{temim2015}. 
To fit the source spectra from each extraction region, we first subtracted the corresponding blank-sky spectrum. 
Both regions were fitted by a power-law model describing the synchrotron emission from the PWN and a non-equilibrium ionization thermal model \texttt{xsvnei} with solar abundances of \citet{wilms2000} to characterize any thermal emission from the SNR. The results for each spectral region are fully consistent with the findings of prior works \citep[e.g.,][see Table~2 for the ``diffuse'' and ``relic'' regions which roughly correspond to the regions employed here]{temim2015}. The results of the fits are listed in Table~\ref{tab:xrayparams}. The spectra and best-fit models for the regions are shown in Figure~\ref{fig:xray_spectra}.

It is apparent from the softer X-ray spectrum that the body is entering the synchrotron cutoff regime, but the X-ray spectrum observed from the ``head'' is much harder. This distinction indicates that the two particle populations are under different external conditions, such as different magnetic field strengths and structures that ultimately change the evolutionary history of each population. The average X-ray spectrum \citep{temim2015} for the entire PWN demonstrates that there are important contributions from both populations to the total X-ray spectrum. We explore the particle properties from the spatially separated radio and X-ray spectra in Section~\ref{sec:model}.

\begin{table*}
\begin{minipage}[b]{0.5\linewidth}
\centering
\begin{tabular}{|c c c |}
\hline
\ Component & Parameter & Best-Fit Value \\
\hline
\ Body or relic 
 & Reduced $\chi^2$ & 1.06 \\
\hline
\ \texttt{tbabs} & $n_H (10^{22})$\,cm$^{-2}$ & 2.30$_{-0.10}^{+0.10}$ \\
\ \texttt{powlaw} (PWN) & $\Gamma$ & 2.43$_{-0.05}^{+0.04}$ \\
\ & Amplitude & 1.60$_{-0.20}^{+0.10}\times10^{-3}$\\
\hline
\ \texttt{vnei} (SNR) & $kT$(keV) & 0.22$_{-0.02}^{+0.04}$ \\
\ & $\tau$ ($10^{11}\,$cm$^{-3}$s) & 2.00$_{-1.00}^{+2.00}$ \\
\ & Normalization & 0.07$_{-0.02}^{+0.08}$ \\
\hline
\end{tabular}
\end{minipage}
\begin{minipage}[b]{0.5\linewidth}
\centering
\begin{tabular}{|c c c |}
\hline
\ Component & Parameter & Best-Fit Value \\
\hline
\ Head or finger 
 & Reduced $\chi^2$ & 1.10 \\
\hline
\ \texttt{tbabs} & $n_H (10^{22})$\,cm$^{-2}$ & 1.90$_{-0.10}^{+0.10}$ \\
\ \texttt{powlaw} (PWN) & $\Gamma$ & 2.09$_{-0.04}^{+0.04}$ \\
\ & Amplitude & 2.50$_{-0.10}^{+0.20}\times10^{-3}$\\
\hline
\ \texttt{vnei} (SNR) & $kT$(keV) & 0.27$_{-0.04}^{+0.04}$ \\
\ & $\tau$ ($10^{11}\,$cm$^{-3}$s) & 1.00$_{-0.60}^{+1.00}$ \\
\ & Normalization & 0.04$_{-0.01}^{+0.05}$ \\
\hline
\end{tabular}
\end{minipage}
\caption{Summary of the 90\% C.L. statistics and parameters for the best-fit model of each component measured in the X-ray analysis.}
\label{tab:xrayparams}
\end{table*}

\subsection{$\gamma$-ray}\label{sec:fermethod}



\subsubsection{\fermi}
The {\it Fermi} Gamma-ray Space Telescope houses both the Gamma-ray Burst Monitor \citep[GBM,][]{meegan2009} and the Large Area Telescope \citep[LAT,][]{atwood2009}. The LAT instrument is sensitive to $\gamma$-rays with energies 50\,MeV to 2\,TeV \citep{atwood2009} and has been continuously surveying the entire sky every 3\,hours since it began operation in August 2008. The latest upgrade to the event reconstruction process and instrumental response functions \citep[dubbed Pass 8,][]{atwood2013} has improved the effective area, point spread function, and background rejection capabilities. These upgrades have enabled improvements for the effective energy and angular resolution. The point spread function (PSF) is energy-dependent, becoming large at lower energies but improving at increasing energies. The PSF reaches optimal resolution at $\gtrsim10$\,GeV with a PSF $\sim0.1$\,$\degree$ (68\% containment). A major advantage to the Pass 8 event reconstruction upgrade is the classification of detected photon events based on the quality of the angular reconstruction. The events are divided into four categories: PSF0, PSF1, PSF2, or PSF3 types. PSF0 has the worst reconstruction quality and each type increases in reconstruction quality such that PSF3 contains the highest-quality events. For the analysis described in this report, we consider all events.

\subsubsection{\fermi Data Analysis}
We re-analyze the \fermi data using an optimized technique in order to maximize the source detection and to most accurately measure the MeV--GeV $\gamma$-ray spectrum. We perform a binned likelihood analysis using 11.5\,years (from 2008 August to 2020 January) of \texttt{P8R3\_SOURCE\_V3} photons with energy between 300\,MeV--2\,TeV from all events. We utilize the latest Fermitools package\footnote{\url{https://fermi.gsfc.nasa.gov/ssc/data/analysis/software/}.} (v.2.0.8) and FermiPy Python 3 package \citep[v.1.0.1,][]{fermipy2017} to perform data reduction and analysis. 
We organize the events by PSF type, as described above, using \texttt{evtype=4,8,16,32} to represent \texttt{PSF0, PSF1, PSF2}, and \texttt{PSF3} components. A binned likelihood analysis is performed on each event type and then combined into a global likelihood function for the ROI to represent all events\footnote{See FermiPy documentation for details: \url{https://fermipy.readthedocs.io/en/0.6.8/config.html}}. Photons detected at zenith angles larger than 100\,$\degree$ were excluded to limit the contamination from $\gamma$-rays generated by CR interactions in the upper layers of Earth's atmosphere. The data were additionally filtered to remove time periods when the instrument was not online (e.g., when flying through the South Atlantic Anomaly).  The $\gamma$-ray data are modeled with the latest comprehensive \fermi source catalog, 4FGL-DR2 \citep{4fgl-dr2}, the LAT extended source template archive for the 4FGL catalog\footnote{\url{https://fermi.gsfc.nasa.gov/ssc/data/access/lat/10yr_catalog}}, and the latest Galactic diffuse and isotropic diffuse templates (\texttt{gll\_iem\_v07.fits} and \texttt{iso\_P8R3\_SOURCE\_V3\_v1.txt}, respectively)\footnote{LAT background models and appropriate instrument response functions: \url{https://fermi.gsfc.nasa.gov/ssc/data/access/lat/BackgroundModels.html}.}. 

We fit the 10$\degree$ region of interest (ROI) with 4FGL point sources and extended sources that are within 15$\degree$ of the ROI center along with the diffuse Galactic and isotropic emission backgrounds. We allow the background components and sources with TS $\geq 25$ and a distance from the ROI center (chosen to be the TeV PWN position) $\leq3.0$\,$\degree$ to vary in spectral index and normalization.

We computed a series of diagnostic test statistic (TS) and count maps in order to search for and understand any residual $\gamma$-ray emission. The TS is defined to be the natural logarithm of the difference in the likelihood of one hypothesis (e.g. presence of one additional source) and the likelihood for the null hypothesis (e.g. absence of source):
\begin{equation}
    TS = 2 \times \log\big({\frac{{\mathcal{L}_{1}}}{{\mathcal{L}_{0}}}}\big)
\end{equation}
The TS quantifies how significantly a source is detected at a given location for a specified set of spectral parameters and the significance of such a detection can be estimated by taking the square root of the TS value for 1 degree of freedom \citep[DOF,][]{mattox1996}. TS values $>25$ correspond to a detection significance $> 4 \sigma$ for 4 DOF.

We generated the count and TS maps in the following energy ranges: 300\,MeV--2\,TeV, 1--10\,GeV, 10--100\,GeV, and 100\,GeV--2\,TeV. The motivation for increasing energy cuts stems from the improving PSF of the \fermi instrument with increasing energies\footnote{See \url{https://www.slac.stanford.edu/exp/glast/groups/canda/lat_Performance.htm} for a review on the dependence of PSF with energy for Pass 8 data.}. We inspect the maps for additional sources, promptly finding a faint point-like $\gamma$-ray source coincident with SNR~G327.1--1.1 in a relatively uncrowded region with no known 4FGL counterpart nearby.\footnote{The closest 4FGL sources are more than 1$\degree$ away in 4FGL--DR2.}

\begin{figure*}
\begin{minipage}[b]{.5\textwidth}
\centering
\includegraphics[width=1.0\linewidth]{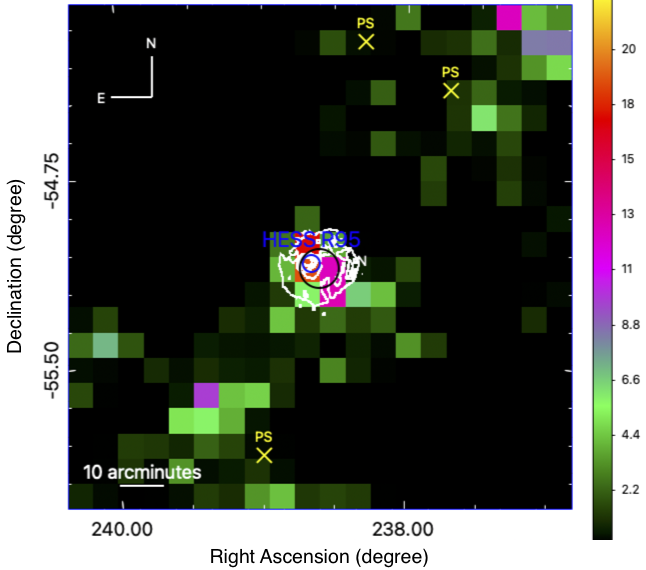}
\end{minipage}
\begin{minipage}[b]{.5\textwidth}
\centering
\includegraphics[width=1.03\linewidth]{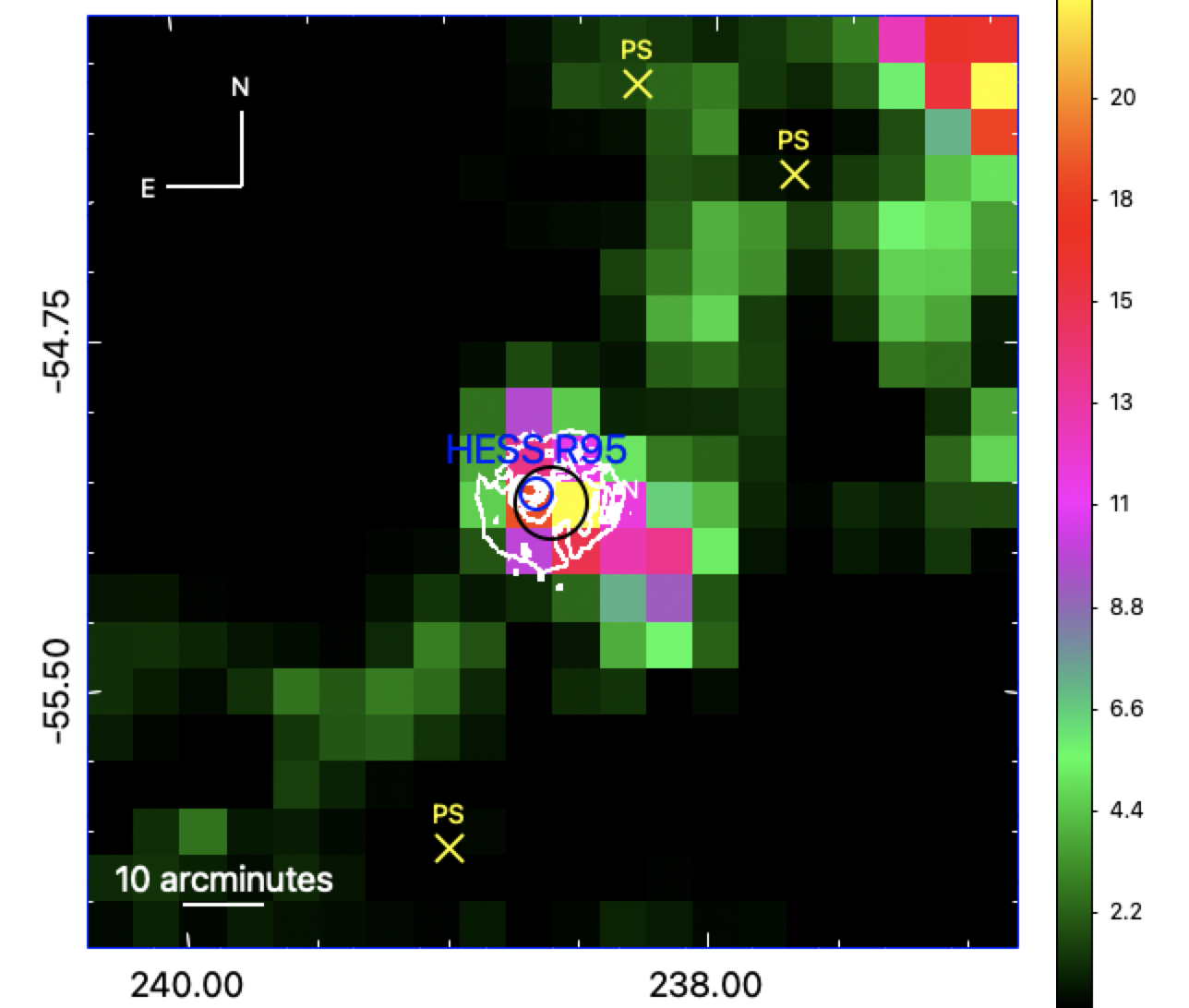}
\end{minipage}
\caption{{\it Left:} 2\,$\degree \times 2$\,$\degree$ TS map of all events from 300\,MeV--2\,TeV. {\it Right:} 2\,$\degree \times 2$\,$\degree$ 1--10\,GeV TS map of all events. The maximum TS at the PWN/SNR position is $\sim$18 and $\sim$23 in the 300\,MeV--2TeV and 1--10\,GeV energy ranges, respectively. The source model used only excludes the point source modelling emission at the location of G327.1--1.1. 
The 843\,MHz (white) radio contours and H.E.S.S. (blue) and \fermi (black) 95\% uncertainty regions are plotted in both panels. }\label{fig:ts_maps}
\end{figure*}

TS maps between energies 300\,MeV--2\,TeV and 1--10\,GeV are shown in Figure \ref{fig:ts_maps} where 
significant residual emission coincides with the position and size of SNR~G327.1--1.1, with a maximum TS value $\sim 18$ in the 300\,MeV--2\,TeV energy range and $\text{TS} \sim 23$ between 1--10\,GeV. It is apparent the residual emission is probably point-like, with little evidence for extension beyond the PSF of the \fermi. 


\begingroup
\begin{table*}
\hspace{-2.5cm}
\scalebox{0.87}{
\begin{tabular}{cccccccc}
\hline
\hline
\ Spectral Model & $\log{L}$ & $\Gamma$ & $\alpha$ or $\Gamma_1$ & $\beta$ or $\Gamma_2$ & $N_0$ (MeV$^{-1}$ cm$^{-2}$ s$^{-1}$) & $E_b$ (GeV) or $a$ & TS \
\\
\hline
Power Law & 2143243.14 & $2.45\pm0.13$ & -- & -- & $(8.12\pm 1.70) \times10^{-13}$ &  -- & 29.05 \\
Log Parabola & 2143245.52  & -- & $2.29\pm0.19$ & $0.20\pm2.85\times10^{-4}$ & $(1.86\pm 0.35) \times10^{-13}$ & 2.00 & 32.95  \\
Power Law with Super Exp. cutoff & 2143245.54 & -- & $1.04\pm1.58$ & $0.67$ (fixed) & $(2.19\pm 2.88) \times10^{-12}$ & $0.01\pm 2.79 \times 10^{-5}$ & 32.97 \\
\hline
\hline
\end{tabular}}
\caption{Summary of the best-fit parameters and the associated statistics with 1-$\sigma$ statistical uncertainties for each of the \fermi point source tests in the 300\,MeV--2\,TeV energy band.}
\label{tab:gamma_spec}
\end{table*}
\endgroup


\subsubsection{\fermi Data Analysis Results}\label{sec:fermi-results}
To model the $\gamma$-ray emission coincident to the PWN inside SNR~G327.1--1.1, we add a point source to the 300\,MeV--2\,TeV source model in addition to known 4FGL sources and the background emission templates for the diffuse Galactic and isotropic components. We first fit the point source with a power law spectrum\footnote{For a review of \fermi source spectral models see \url{https://fermi.gsfc.nasa.gov/ssc/data/analysis/scitools/source_models.html}.},
\begin{equation}
    \frac{dN}{dE} = N_{0} \big(\frac{E}{E_0}\big)^{-\Gamma}
\end{equation}
and photon index $\Gamma=2$. We localize the point source using \texttt{GTAnalysis.localize} to find the best-fit position is at (R.A., Dec.) = (238.595$\degree$, --55.110$\degree$) (J2000), an offset of 0.06 $\degree$ from the exact PWN location, and a 95\% positional uncertainty radius $r=0.079 \degree$, which overlaps well with the radio and X-ray positions of the PWN. With the new position, we then allow the spectral index and normalization to vary. The TS for a localized point source with a power law spectrum and photon index $\Gamma=2.45\pm0.13$ is $29.05$.

\begin{figure}[!t]
\hspace{-.5cm}
\includegraphics[width=1.1\linewidth]{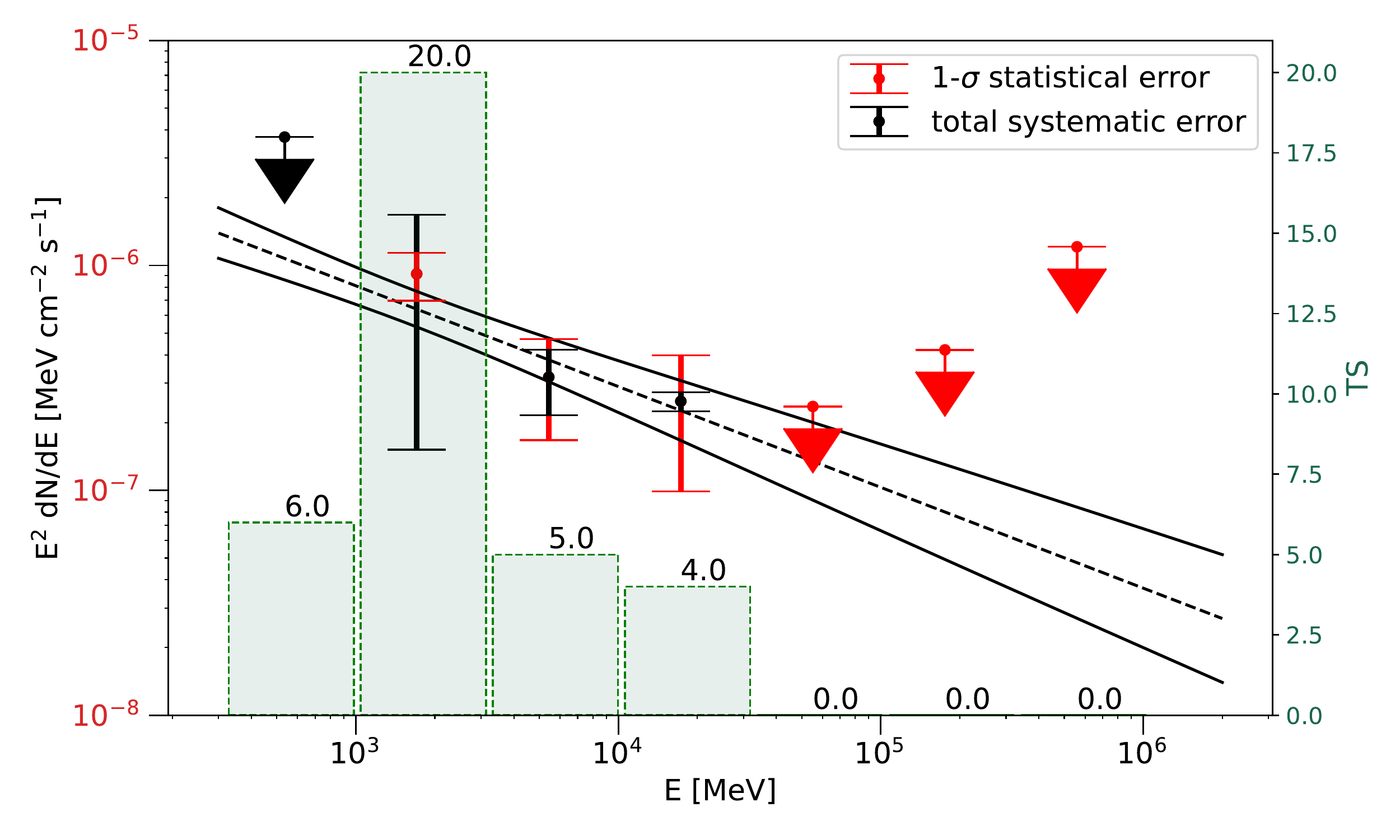}
\caption{The best-fit $\gamma$-ray SED between 300\,MeV and 1\,TeV over 7 equally-spaced logarithmic energy bins with 1-$\sigma$ statistical uncertainties (red) and the quadratic sum of systematic errors (black, see Section \ref{sec:sys_err}). TS values for each spectral bin are plotted as the green histogram. The best-fit power-law spectrum is fit with a photon index $\Gamma_\gamma = 2.45\pm 0.13$.}\label{fig:gamma_sed}
\vspace{0.05cm}
\end{figure}

We investigate the spectral properties of the $\gamma$-ray emission by changing the spectrum to a log parabola representation:
\begin{equation}
    \frac{dN}{dE} = N_{0} \big(\frac{E}{E_b}\big)^{-(\alpha+\beta\log{E/E_b})}
\end{equation}
We repeat the same method a third time, replacing the log parabola point source with a point source described by a power law with a super exponential cutoff (PLEC),
\begin{equation}
    \frac{dN}{dE} = N_{0} \big(\frac{E}{E_0}\big)^{-\Gamma_1} \exp{\big(-a E^{\Gamma_2}\big)}
\end{equation}
See Table \ref{tab:gamma_spec} for a summary of the best-fit parameters for each point source test. Given that all three spectral models provide similar statistical fits, we model the emission with the power-law spectrum since it is the simplest and has the fewest degrees of freedom. As shown in the $\gamma$-ray SED in Figure~\ref{fig:gamma_sed}, almost all of the emission is observed between 1--10\,GeV.

\begingroup
\begin{table*}
\centering
\scalebox{0.87}{
\begin{tabular}{cccccccc}
\hline
\hline
\ Spatial Template & TS & TS$_{ext}$ & R.A., Dec. (J2000) & Best-fit radius or sigma ($\degree$) & 95\% radius upper limit ($\degree$) \
\\
\hline
Point Source & 29.05 & -- & 238.59, --55.11 & -- & -- \\
Radial Disk & 29.29 & 0.25 & 238.65, --55.07 & 0.004 & 0.083 \\
Radial Gaussian & 29.30 & 0.26 & 238.65, --55.07 & 0.003 & 0.077 \\
\hline
\hline
\end{tabular}}
\caption{Summary of the best-fit parameters and the associated statistics for each spatial template tested in the 300\,MeV--2\,TeV energy band.}
\label{tab:extent}
\end{table*}
\endgroup


We run extension tests for the best-fit point source in FermiPy utilizing \texttt{GTAnalysis.extension} and the two spatial templates supported in the FermiPy framework, the radial disk and radial Gaussian. Both of these extended templates assume a symmetric 2D shape with width parameters radius and sigma, respectively. We allow the position to vary when finding the best-fit spatial extension for both templates. The summary of the best-fit parameters for the extended templates are listed in Table \ref{tab:extent}. The $\gamma$-ray emission shows no evidence for extension.

The origin of the emission is unlikely to be the SNR, given the distinct location inside the SNR shell in addition to a general lack of evidence for an energetic interaction between the SNR and the ISM that would produce $\gamma$-rays. The 300\,MeV--2\,TeV energy flux $F_{\gamma} = 5.12 (\pm 0.97) \times 10^{-12}$\,ergs cm$^{-2}$ s$^{-1}$ gives a $\gamma$-ray luminosity $L_\gamma = 4.94 \times 10 ^{34} \big(\frac{d}{9kpc}\big)^2$\, ergs s$^{-1}$. Taking the predicted current spin-down power $\dot{E} \sim 3.1 \times 10^{36}$\,ergs s$^{-1}$ of the pulsar from \citet{temim2015}, 
we measure the $\gamma$-ray conversion efficiency $\eta_\gamma \sim 0.016$, which is comparable to other GeV PWNe \citep{acero2013}. However, we cannot exclude a pulsar contribution, as several $\gamma$-ray pulsars have similar values of $\eta$ \citep{2pc}. The $\gamma$-ray emission is not bright enough to perform a reliable pulsation search and we cannot rule out the scenario from the $\gamma$-ray SED alone. The results reported here are consistent with those reported in \citet{xiang2021}, where the coincident $\gamma$-ray emission is detected just below the Fermi--LAT threshold of $4\,\sigma$, $\text{TS} = 22.94$ for 4 DOF and characterized best with photon index $\Gamma = 2.35 \pm 0.24$.

\subsubsection{Systematic Error}\label{sec:sys_err}
We account for systematic uncertainties introduced by the choice of the interstellar emission model (IEM) and the effective area, which mainly affect the spectrum of the faint $\gamma$-ray emission. We have followed the prescription developed in \citet{depalma2013,acero2016}, which generated eight alternative IEMs using a different approach than the standard IEM \citep[see][for details]{acero2016}. For this analysis, we employ the eight alternative IEMs (aIEMs) that were generated for use on Pass 8 data in the \fermi Galactic Extended Source Catalog \citep[FGES,][]{ackermann2017}. The $\gamma$-ray point source coincident with PWN~G327.1--1.1 is refit with each aIEM to obtain a set of eight values for the spectral flux that we compare to the standard model following equation (5) in \citep{acero2016}. We estimate the systematic uncertainties introduced by the instrument response function (IRF) uncertainties\footnote{\url{https://fermi.gsfc.nasa.gov/ssc/data/analysis/LAT_caveats.html}} while enabling energy dispersion as follows: $\pm 3\%$ for $E < 100$\,GeV, $\pm 4.5 \%$ for $E = 175\,$GeV, and $\pm 8\%$ for $E = 556\,$GeV. Since the IEM and IRF systematic errors are taken to be independent, we can evaluate both and perform the quadratic sum for the total systematic error. We find that the systematic errors are most important in the first two energy bins of the $\gamma$-ray spectrum, which effectively convert the first data point to an upper limit, but are negligible in the remainder of energy bins.   The corresponding quadratic sum of the systematic errors in the first four energy bins are plotted in Figure~\ref{fig:gamma_sed} in black.

\section{Broadband Modeling}\label{sec:model}
\begin{figure*}
\hspace{-.4cm}
\centering
\includegraphics[width=0.9\linewidth]{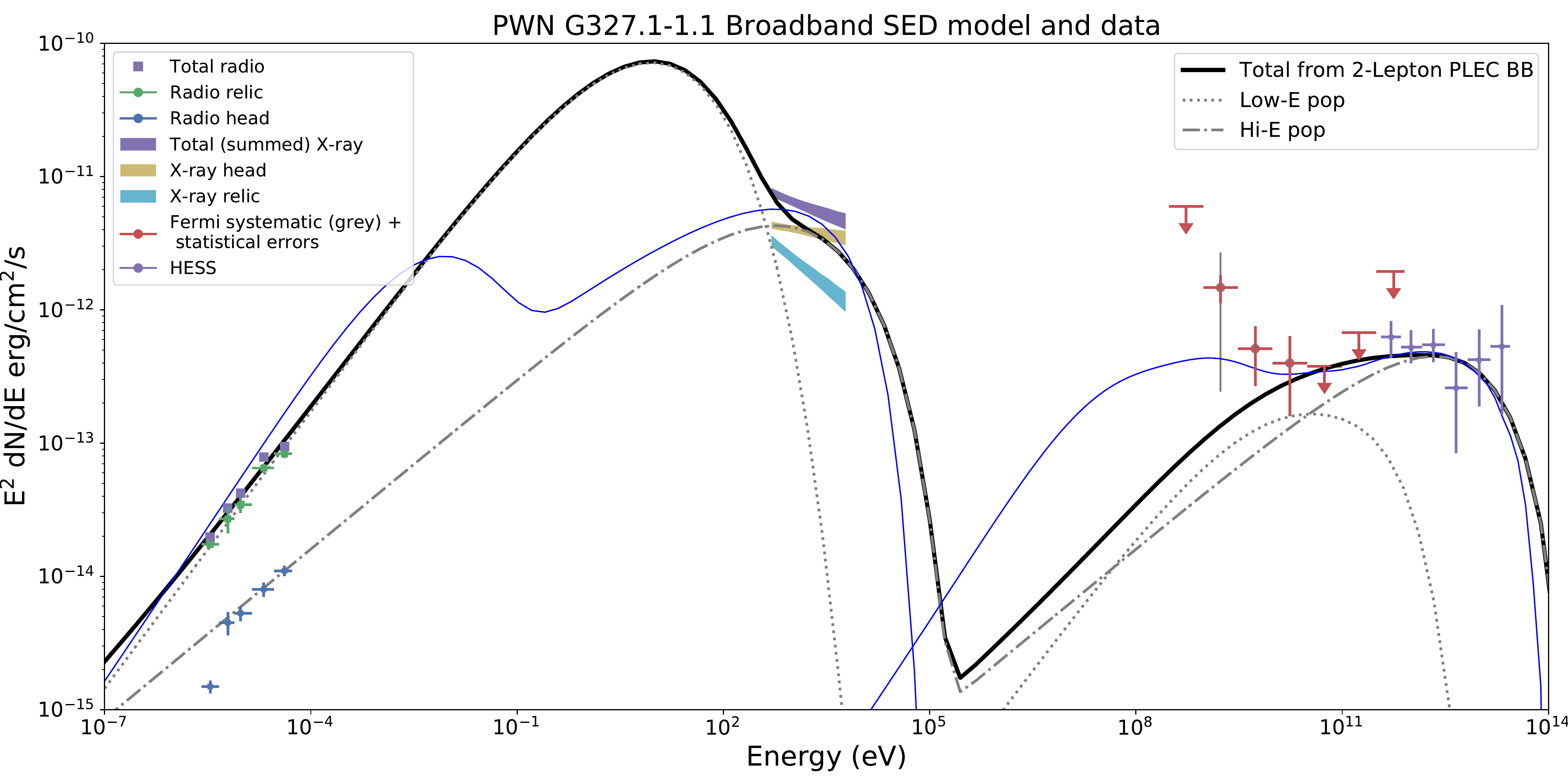}
\caption{Best-fit broadband model fits for emission from the evolved PWN in G327.1--1.1 along with radio observations of the PWN \citep[][]{ma2016}, X-ray fluxes and spectral indices (this work), new \fermi data (this work), and H.E.S.S. $\gamma$-ray emission \citep[][]{acero2011}. For comparison, we plot the broadband representation derived from the evolved BPL injection spectrum in \citet{temim2015} in solid blue. 
}\label{fig:broadband_sed}
\end{figure*}

A broadband model was derived for G327.1--1.1 from a semi-analytic simulation that predicts the evolution of a PWN inside a nonradiative SNR and was reported in \citet{temim2015}. Assuming a broken power law (BPL) particle injection spectrum and given a set of input parameters, the model simultaneously predicts the PWN and SNR size, the PWN magnetic field, and the PWN and SNR expansion velocities as a function of age. The final particle and photon spectra are measured by evolving the BPL injection spectrum with time while accounting for the evolution of the PWN and host SNR. The particle spectrum evolves until it corresponds to the age of the SNR at which the simulated PWN and SNR radii match the observed sizes \citep[see][for details]{gelfand2009,temim2015}. The corresponding broadband photon spectrum can reproduce well the radio, X-ray, and $\gamma$-ray data of G327.1--1.1 considered by \citet{temim2015}, and suggests the remnant is old $\tau \sim 18,000\,$yrs with distinct particle components apparent in the simulated SED (see Figure~\ref{fig:broadband_sed}).

The resolved PWN morphology in radio and X-ray gives a unique opportunity to study the individual properties of the particle populations through their distinct broadband spectral features, first investigated by \citet{temim2015}. In this section we attempt to model the particle populations independently in an effort to characterize each component that contributes to the observed broadband emission, incorporating the new \fermi data, and compare to the model presented in \citet{temim2015}.
We use the NAIMA Python package \citep{naima}, designed to compute the non-thermal radiation from relativistic particle populations. NAIMA is a simple radiation code, in contrast to the semi-analytic method employed by \citet{temim2015}, that uses a constant particle distribution and hence it does not account for the evolution of the total particle energy and spectrum as the PWN and host SNR evolve. 
To explore the possible particle distributions, we assume a power law shape with an exponential cutoff (PLEC),
\begin{equation}
f(E) = A \big(\frac{E}{E_0}\big)^{-\Gamma} 
\exp\big({-{\frac{E}{E_{c}}}\big)} \\
\end{equation}
where $A$ is in eV$^{-1}$. 
We then input similar values as those estimated in \citet{temim2015}, such as the particle index and energy break $E_{break}$ in the particle spectrum, and also consider the same three photon fields for the Inverse Compton Scattering (ICS) component. The photon fields consist of the cosmic microwave background (CMB) with energy density $\Omega_{CMB} = 0.261$\,eV cm$^{-3}$ and temperature of 2.72\,K, and two warmer photon fields from heated dust and nearby starlight, with energy densities 1.5\,$\Omega_{CMB}$ and 2.0\,$\Omega_{CMB}$ and temperatures $T=3.5\times10^{3}\,$K and $T = 50\,$K, respectively.  
We then test the combination of free PLEC parameters that can best explain the broadband spectrum. 

While a single particle distribution can generally reproduce the observed synchrotron and IC spectral shapes, it cannot reproduce the radio spectrum nor explain the MeV--GeV $\gamma$-rays adequately. The sum of two leptonic populations under the same ambient photon fields thus represents the broadband spectrum of PWN~G327.1--1.1: one component belonging to the older, larger and displaced radio relic and the second component belonging to the compact, younger X-ray nebula. The observed displacement of the radio and X-ray structures, in addition to the estimated age for the nebula as a whole, strongly indicate the presence of distinct particle components under different physical conditions as a result of the asymmetric crushing of the PWN. As such, we fit the magnetic field associated with each leptonic population independently. This assumption is further supported by the distinct radio and X-ray spectra measured for each population. As shown in Figure~\ref{fig:broadband_sed}, the lower-energy population clearly dominates the radio spectrum for the nebula, but is in the synchrotron cut-off regime in the X-ray band, while the higher-energy population is faint in radio but bright in X-ray. Observationally, the larger radio to $\gamma$-ray ratio in the relic implies that the magnetic field is larger than in the X-ray head.


\begingroup
\begin{table*}
\centering
\begin{tabular}{ccccccc}
\hline
\hline
\ Particle Spectrum & $\Gamma_1$ & $\Gamma_2$ & $E_{c}$ or $E_{break}$ (TeV) & $B$ ($\mu$G) \
\\
\hline
PLEC1 (low-energy pop.) & 1.61 & -- & 0.99 & 101.00 \\
PLEC2 (high-energy pop.) & -- & 2.15 & 41.69 & 10.76 \\
BPL \citep{temim2015} & 1.48 & 2.20 & 0.31 & 10.76 \\
\hline
\hline
\end{tabular}
\caption{Summary of the best-fit parameters of the broadband models in Figure~\ref{fig:broadband_sed} with those of \citet{temim2015} for comparison. }
\label{tab:broadband_params}
\end{table*}
\endgroup

Table~\ref{tab:broadband_params} lists the best-fit parameters for the two-leptonic PLEC particle spectra investigated here along with the best-fit parameters from the evolved BPL particle spectrum in \citet{temim2015} for comparison. The corresponding best-fit broadband spectrum for the two PLEC lepton populations is displayed in Figure~\ref{fig:broadband_sed}. We find the low-energy (i.e., older) population is best characterized by a particle index before the energy cutoff $\Gamma = 1.61$. We find the high-energy (i.e., younger) population is best-fit with a particle index before the cutoff $\Gamma = 2.15$. The implied magnetic field is much higher for the lower-energy population $B = 101\,\mu$G, about ten times higher than the higher-energy population $B = 10.8\,\mu$G. 
The low-energy cutoff is $E_{c}=0.99\,$TeV and the high-energy cutoff is $E_{c}=41.69\,$TeV.

The low- and high-energy particle indices reported here are similar to the low- and high-energy particle indices 1.48 and 2.20 found in \citet{temim2015} which were motivated by the observed average radio and X-ray indices. Because the low-energy population clearly dominates the radio emission, the comparable low-energy index is not surprising. The X-ray emission is a combination of the two populations as shown in our Figure~\ref{fig:broadband_sed} such that the similarity for the high-energy index found here and in \citet{temim2015} is expected. Moreover, the X-ray spectrum of each particle component shows two important distinctions: the older population is entering the synchrotron cutoff regime while the younger population does not show strong evidence for synchrotron cutoff in the 0.5--6\,keV X-ray band. This suggests the highest-energy particles are not suffering the same synchrotron losses as the lowest-energy particles and is supported by the high magnetic field value inferred from the lower-energy population. The high-energy magnetic field component is much lower, but matches the magnetic field strength inferred from \citet{temim2015}. 

It is clear that the X-ray spectrum from the relic is underpredicted by the low-energy population in the broadband model. This could be explained by the assumption that both particle populations are characterized by a power law with an exponential cutoff distribution. The particle spectrum of an evolved PWN such as G327.1--1.1 has been shown to deviate beyond a simple power-law distribution even if the initial injection spectrum is a power-law \citep{gelfand2009}, such that if the low-energy population truly represents the relic nebula, the particle spectrum may not be best described in power-law form. This limitation is not present in the \citet{temim2015} model, however, since the particle spectrum evolution is considered. The synchrotron emission peaks at a flux about an order of magnitude higher than what is predicted in the \citet{temim2015} model and is shifted to lower energy by roughly one order of magnitude. This is due to the much stronger magnetic field strength combined with the low-energy cutoff that characterizes the oldest particle population. The last most striking difference is the \citet{temim2015} model can accurately predict the MeV--GeV $\gamma$-ray spectrum, intersecting with all of the most constraining data points in the \fermi SED. The IC emission in the \citet{temim2015} model has a two-peak spectral shape that manifests from the evolutionary history of the particles throughout the PWN's lifetime, which accounts for the total energy input of the PWN decreasing with time once the pulsar exits the nebula. 

The divergences in the model fits, especially for the oldest particles, demonstrate the importance in considering the evolutionary history for broadband studies of PWNe. 
Despite the use of simplified radiative models to develop a new broadband representation,  both models depict a physical scenario where the reverse shock has compressed the relic PWN while the exiting pulsar generates a new X-ray nebula which may have minimal to no impact from the reverse shock and subsequent compression. We discuss the physical implications of the models in the following section.

\section{Discussion}\label{sec:discuss}

We have presented a broadband model for the PWN emission in G327.1--1.1 that is characterized by the sum of two-lepton PLEC spectra and compared to the evolved particle spectrum derived in \citet{temim2015}. The best-fit model estimated from the evolved particle spectrum is based on the dynamical model for PWN evolution inside a non-radiative supernova remnant, first described in \citet{gelfand2009}. We further investigate the physical implications of the presented model and of the \citet{temim2015} model by comparing the predicted observable properties from the evolutionary phases expected for PWNe from \citet{gelfand2009} with those observed from the PWN inside SNR~G327.1--1.1. 

The initial expansion stage is the first phase of PWN evolution and occurs from the PWN pressure being much greater than the pressure of the surrounding SN ejecta. 
Throughout the PWN expansion, the pulsar radiates more energy than the PWN loses from adiabatic expansion and synchrotron losses. The expansion phase introduces a rapid decline in the PWN magnetic field, which results in a steep decline in synchrotron luminosity with time. As a result, adiabatic losses will dominate over radiative losses until the reverse shock crushes the PWN, which also marks the end of the first expansion stage.  
The particle spectrum is dominated by previously injected particles at all energies for the majority of the PWN expansion. 
Similarly, the corresponding photon spectrum is dominated by previously injected particles at all energies, forming a single peak structure for both synchrotron and Inverse Compton emission. The initial expansion stage occurs within the first $\sim 1-5$\,kyrs of the PWN lifetime, but depends on the ambient ISM density and the host SNR evolution. The initial expansion phase ends when the SNR reverse shock first collides with the PWN.

The particle spectrum and consequent photon spectrum during the first contraction phase will see dramatic changes once collision with the reverse shock has begun. The synchrotron lifetime of the highest energy particles will become significantly less than the PWN age and will introduce an energy break in the particle spectrum. The most recently injected particles dominate beyond this energy break. During the PWN compression, the synchrotron luminosity will increase due to the strengthening magnetic field. Meanwhile, the IC luminosity depends mostly on the highest-energy particles, which are those most recently injected by the central pulsar.
The corresponding photon spectrum sees the synchrotron emission peak decrease in energy due to the overall energy decreasing with time from synchrotron losses. The IC emission peak also expects to decline in the first contraction stage both in part due to the central pulsar input and to the short synchrotron lifetime of the particles, such that only the highest energy particles contribute to the TeV emission. 


After the PWN undergoes significant compression, the maximum particle energy and magnetic field strength will eventually begin to decrease. The corresponding photon spectrum will see another shift in the synchrotron and IC peaks such that the majority of the broadband emission occurs in the radio, soft and hard X-ray bands. The continuous injection of high-energy particles by the central pulsar will result in two distinct components visible in both the particle and photon spectra: a lower energy population composed of particles injected at earlier times and a higher energy component composed of recently injected particles. 
The corresponding photon spectrum features the double-peaked structure in both synchrotron and IC emission processes, akin to the \citet{temim2015} model for G327.1--1.1 and similar to the one presented in Section~\ref{sec:model}. At this stage, the radio emission is attributed to synchrotron radiation from the oldest particles and the GeV emission from the same population ICS against ambient photons, while the rest of the emission is dominated by recently injected particles. 
In the case where the pulsar exits the nebula and does not re-enter, the pulsar still continues to inject high-energy particles into its surroundings, generating a new high-energy PWN that is separate from the relic PWN it has left behind. 

Much of the depicted evolutionary cycle depends on the unique physical properties of the system and surroundings such as the progenitor characteristics, the ISM density, and the pulsar spin-down. The observed morphology and spectral characteristics of G327.1--1.1 suggest it to be an evolved system that has already encountered or is currently encountering the passage of the reverse shock through the SNR interior, crushing the PWN in its first contraction. 

The evolutionary cycle described here is only valid for one-dimensional, spherically symmetric, non-radiative host SNR systems. The spherically symmetric condition introduces a series of compression and re-expansion cycles to the nebula that are initiated by the passage of the reverse shock. For increasingly asymmetric systems such as G327.1--1.1, the evolutionary cycle after first compression can differ significantly \citep{gelfand2009,temim2015,bandiera2020}. 
The thin-shell approximation is also used in the model presented by \citet{temim2015}, where the PWN is assumed to be surrounded by a thin shell of swept-up SN ejecta.  If this assumption is not realistic, the impact of the reverse shock and subsequent compression can be over-predicted \citep[e.g.,][]{bandiera2020,fiori2022}. 
Despite these limitations, particularly in characterizing the initial compression stage, the physical implications from the simplified model we present in Section~\ref{sec:model} and the one reported in \citet{temim2015} are consistent with the observational properties for the system. 

The size of the SNR ($\sim 22$\,pc) at the estimated distance $\sim 9$\,kpc and the displaced PWN radio and X-ray counterparts support the scenario of an older system that has experienced an interaction with the SNR reverse shock. It is likely the reverse shock has displaced the PWN to the East and, combined with the velocity of the presumed pulsar \citep{temim2009,temim2015}, has enabled the pulsar to exit the PWN leaving behind a radio relic PWN while continuously injecting high-energy particles into the surroundings, generating the X-ray nebula in its wake. 
The radio and X-ray spectra further validate the scenario depicted here, where significant synchrotron losses are apparent, especially for the oldest particles in the relic, and likely being the direct result of the reverse shock interaction. The observed morphology also implies that the highest-energy emission, which is observed close to the putative pulsar, is near the maximum particle energy, and indicates a steepening in the particle spectrum \citep{temim2015}. 


\section{Conclusions}\label{sec:conclude}
We have reported the detection of faint $\gamma$-ray emission discovered coincident in location with the evolved SNR~G327.1--1.1. The position and extent is consistent with an origin from the high-energy PWN which is also known to emit TeV $\gamma$-rays as HESS~J1554--550. We perform a multiwavelength investigation to explore the broadband models supported by observations, utilizing a one-dimensional radiative modeling package \texttt{NAIMA}. We compare the results to the in-depth semi-analytic simulation that predicts the evolved particle injection spectrum based on hydrodynamically derived quantities such as supernova explosion energy and ejecta mass and reported in \citet{temim2015}. We investigate the broadband emission structure based on our understanding of the PWN evolutionary sequence outlined in \citet{gelfand2009}. We find that the broadband model presented here and the one reported in \citet{temim2015}, although obtained through very different techniques, still provide consistent and plausible physical implications.
The faint $\gamma$-ray emission reported here may have a pulsar contribution, particularly for energies $E < 10$\,GeV. Comparison of the two broadband models explored here also suggests that particle injection history is required for accurate $\gamma$-ray emission models of an evolved system like G327.1--1.1. The Cherenkov Telescope Array (CTA)\footnote{\url{https://www.cta-observatory.org/}} will be $\sim 5-10$ times more sensitive than H.E.S.S. at 1\,TeV with an angular resolution that could spatially resolve the PWN TeV morphology to be compared to the radio and X-ray components, thus revealing how the distinct particle components contribute to the lower-energy $\gamma$-ray emission.


\begin{acknowledgements}
The \fermi Collaboration acknowledges generous ongoing support from a number of agencies and institutes that have supported both the development and the operation of the LAT as well as scientific data analysis. These include the National Aeronautics and Space Administration and the Department of Energy in the United States, the Commissariat \a'a l'Energie Atomique and the Centre National de la Recherche Scientifique / Institut National de Physique Nucl\a'eaire et de Physique des Particules in France, the Agenzia Spaziale Italiana and the Istituto Nazionale di Fisica Nucleare in Italy, the Ministry of Education, Culture, Sports, Science and Technology (MEXT), High Energy Accelerator Research Organization (KEK) and Japan Aerospace Exploration Agency (JAXA) in Japan, and the K. A. Wallenberg Foundation, the Swedish Research Council and the Swedish National Space Board in Sweden.

Additional support for science analysis during the operations phase is gratefully acknowledged from the Istituto Nazionale di Astrofisica in Italy and the Centre National d'\'Etudes Spatiales in France. This work performed in part under DOE Contract DE- AC02-76SF00515.
\end{acknowledgements}

\software{CIAO \citep[v4.12][]{ciao2006}, FermiPy \citep[v.1.0.1][]{fermipy2017}, Fermitools: Fermi Science Tools \citep[v2.0.8][]{fermitools2019}, NAIMA \citep{naima}}

\bibliographystyle{aa}
\bibliography{aa.bib}
\end{document}